\documentclass[%
 preprint,
superscriptaddress,
 amsmath,amssymb,
 aps,
]{revtex4-1}

\usepackage{graphicx}
\usepackage{dcolumn}
\usepackage{bm}
\usepackage{amsfonts}
\usepackage{hyperref}
\usepackage[mathlines]{lineno}

\newcommand{\snn}{\sqrt{s_\text{NN}}}

\begin{document}

\preprint{}

\title{Neutron density fluctuation and neutron-proton correlation from AMPT model}

\author{Zuman Zhang}\thanks{Email: zuman.zhang@hue.edu.cn}
\affiliation{School of Physics and Mechanical Electrical \& Engineering, Hubei University of Education, Wuhan 430205, China}
 \affiliation{Institute of Theoretical Physics, Hubei University of Education, Wuhan 430205, China}
\affiliation{Key Laboratory of Quark and Lapten Physics (MOE), Central China Normal University, Wuhan 430079, China}
\author{Ning Yu}\thanks{Email: ning.yuchina@gmail.com} 
\affiliation{School of Physics and Mechanical Electrical \& Engineering, Hubei University of Education, Wuhan 430205, China}
 \affiliation{School of Physics \& Electronic Engineering, Xinyang Normal University, Xinyang 464000, People's Republic of China}
 \affiliation{Institute of Theoretical Physics, Hubei University of Education, Wuhan 430205, China}
 \affiliation{Key Laboratory of Quark and Lapten Physics (MOE), Central China Normal University, Wuhan 430079, China}
 \author{Hongge Xu}
 \affiliation{School of Physics and Mechanical Electrical \& Engineering, Hubei University of Education, Wuhan 430205, China}
 \affiliation{Institute of Theoretical Physics, Hubei University of Education, Wuhan 430205, China}


\date{\today}

\begin{abstract}
Using the multiphase transport (AMPT) model, we study the relative neutron density fluctuation and neutron-proton correlation in matter produced by Au+Au collisions at $\snn = $7.7-200 GeV. The rapidity, centrality, and energy dependence of these two observations are also discussed. The light nuclei yield ratio of proton, deuteron, and triton $N_tN_p/N_d^2$ calculated directly from the relative neutron density fluctuation and neutron-proton correlation, decreases with rapidity coverage and increases with collision centrality. Our study also found that the ratio does not exhibit any non-monotonic behavior in collision energy dependence. Since there is no first-order phase transition or critical physics in the AMPT model, our work provides a reference for extracting the relative neutron density fluctuation from light nuclei production in experiments.

\begin{description}\item[PACS numbers]
\verb+25.75.Nq, 24.10.Lx, 24.10.Pa+
\end{description}
\end{abstract}

\maketitle

Quantum Chromodynamics (QCD) is the fundamental theory of the strong interaction. One of the main goals of ultra-relativistic nuclear collisions is to study the properties of the QCD phase diagram~\cite{RN21}. Lattice QCD calculations show that the transition between the hadron gas and the Quark-Gluon Plasma (QGP) is a smooth crossover at zero baryon chemical potential~\cite{RN175}. While in the large $\mu_B$ region, some effective models predicted this transition is of the first order~\cite{RN176}. These two conclusions indicate the existence of a critical point, where the first order phase transition ends. Theoretically, many efforts have been made to determine the critical point in Lattice QCD~\cite{RN177,RN178,RN179,RN180,RN181} and model~\cite{RN182}, but its location or even its existence has not been confirmed. Experimental scientists are systematically studying the phase structure of the QCD matter in high baryon density region. Finding critical point is one of the main objectives of the Beam Energy Scan (BES) program at the Relativistic Heavy Ion Collider (RHIC). This is also the main physics motivation of future accelerators such as Facility for Anti-Proton and Ion Research (FAIR) in Darmstadt and Nuclotron based Ion Collider fAcility (NICA) in Dubna. Experimental confirmation of the existence of the QCD critical point will be a milestone in exploring the properties of the QCD phase structure.

In the vicinity of the QCD critical point, the correlation length of the system and density fluctuations will become large. Fluctuations of conserved quantities, such as net-baryon, net-charge and net-strangeness, are sensitive to the correlation length. The STAR experiment has measured the high-order cumulants and second-order off-diagonal cumulants of net-proton, net-charge, and net-kaon multiplicity distributions~\cite{RN182,RN183,RN184,RN185,RN186,RN187,RN188} in Au+Au collisions at $\snn = $7.7-200 GeV. A non-monotonic behavior as a function of $\snn$ in fourth order net-proton cumulant ratio was observed in the most central collisions with a minimum of around 19.6 GeV. Without the physics of critical point, this behavior cannot be described in model calculations.

Light nuclei production is also predicted to be sensitive to the baryon density fluctuations assuming they are formed from nucleon coalescence~\cite{RN14,RN9,RN118,RN216}. The light nuclei yield ratio $N_pN_t/N_d^2$ of produced proton($p$), deuteron($d$), and triton($t$) can be described by the relative neutron density fluctuation $\sigma_n^2/\langle n\rangle^2$ (also denoted as $\Delta\rho_n$ in Ref.~\cite{RN9}), where $\sigma_n$ and $\langle n\rangle$ are standard deviation and mean value of neutron production. The STAR experiment observed a non-monotonic energy dependence of the light nuclei yield ratio in the most central Au+Au collisions with a maximum peak around $\snn=20\sim30$ GeV~\cite{RN172,RN173,2209.08058}, indicating a large density fluctuation around this energy range. In the coalescence model, it is assumed that the neutron-proton correlation $C_{np}$ disappears or is at least independent of the collision energy and centrality. In this hypothesis, the light nuclei yield ratio can be used to characterize the relative neutron density fluctuation. In this way, we can study the energy dependence of the relative neutron density fluctuation and the location of the critical point or critical region.

To study the $C_{np}$ in heavy ion collisions, a multiphase transport (AMPT) model~\cite{RN47} will be used to analyze the rapidity, collision energy, and centrality dependence of nucleon density fluctuations and $C_{np}$ at $\snn = $7.7, 11.5, 14.5, 19.6, 27, 39, 54.4, 62.4 and 200 GeV. Our results can shed some light on how to extract the neutron density fluctuations from the light nuclei ratio. First, we will give a brief introduction to the AMPT model. Then, the relationship between the relative neutron density fluctuation and the light nuclei yield ratio in heavy ion collisions is given. The behavior of neutron density fluctuation and $C_{np}$ will be given. Finally, we will give the summary.

The AMPT is a hybrid model consisting of four main components, the initial conditions, partonic interactions, conversion from partonic to hadronic matter, and hadronic interactions~\cite{RN47}. The default version of the AMPT involves only minijet partons in the parton cascade and uses the Lund string fragmentation for hadronization~\cite{RN212}. On the other hand, the string melting version of the AMPT model, where all the excited strings are converted to partons and a quark coalescence model is used to describe the hadronization.
Typically, the default version gives a reasonable description of $dN/d\eta$, $dN/dy$, and the $p_T$ spectra, while the string melting version describes the magnitude of the elliptic flow but not the $p_T$ spectra. The string melting version, with a new set of parameters, can well reproduce the $p_T$ spectra and elliptic flows at RHIC top energy~\cite{RN97}. In this paper, the relative nucleon density fluctuation and $C_{np}$ are studied by using this set of parameters.

Base on the Refs.~\cite{RN9,RN172}, the light nuclei are formed through nucleon coalescence. If we ignore the binding energy, the number of deuterons and tritons can be expressed as,
\begin{align}
    N_d&=\frac{3}{2^{1/2}}\left(\frac{2\pi}{mT}\right)^{3/2}N_p\langle\rho_n\rangle(1+C_{np})\\
    N_t&=\frac{3^{3/2}}{4}\left(\frac{2\pi}{mT}\right)^{3}N_p\langle\rho_n\rangle^2(1+\Delta\rho_n+2C_{np}),
\end{align}
in which $N_p$ is the number of protons and $m$ is the nucleon mass, in which we ignore the mass difference between proton and neutron. The three-nucleon correlation $C_{nnp}$ is also neglected in the calculation of triton multiplicity, the effects of which need to be further investigated. $\langle\rho_n\rangle$ is the neutron density. $T$ is the local effective temperature at coalescence. $\Delta\rho_n = \sigma^2_n/\langle\rho_n\rangle^2$ is the dimensionless relative neutron density fluctuation.
\begin{equation}
    C_{np}=\frac{\textrm{Cov}(\rho_n,\rho_p)}{\langle\rho_n\rangle\langle\rho_p\rangle}=\frac{r_{np}\sigma_{n}\sigma_{p}}{\langle\rho_n\rangle\langle\rho_p\rangle}
\end{equation}
is the neutron-proton correlation, where $\textrm{Cov}(\rho_n,\rho_p)$ and $r_{np}$ are the covariance and correlation coefficient between neutron and proton density.

To eliminate the energy dependence on $T$, the light nuclei ratio is considered.
\begin{equation}
    R=\frac{N_tN_p}{N_d^2}=\frac{1}{2\sqrt{3}}\frac{1+\Delta\rho_n+2C_{np}}{(1+C_{np})^2}
    \label{lnr}
\end{equation}

We can find
\begin{align}
    \frac{\partial R}{\partial \Delta\rho_n}&=\frac{1}{2\sqrt{3}}\frac{1}{(1+C_{np})^2}>0\\
    \frac{\partial R}{\partial C_{np}}&=-\frac{\Delta\rho_n+C_{np}}{\sqrt{3}}\frac{1}{(1+C_{np})^3}<0
\end{align}

This ratio increases as the relative neutron density fluctuation $\Delta\rho_n$ and decreases with the correlation $C_{np}$. The $C_{np}$ is very important to extract the true signal of the relative neutron density fluctuation from light nuclei ratio in heavy ion collision.

In AMPT, the event-by-event multiplicity and fluctuation of proton $N_p,S_p$ and neutron $N_n,S_n$ can be extracted from AMPT. In our study, about one million events for each collision energy were generated for the Au + Au collision. The neutron density is calculated as $\rho_n=N_n/V$, in which $V$ is the system volume. The neutron density fluctuation is calculated as $S_{\rho_n}=S_n/V^2=\sigma_n^2/V^2$. Both density and density fluctuation are dependent on system volume. To remove the system volume effect, we use two dimensionless statistical quantities $\sigma_p/\langle\rho_p\rangle=\sqrt{S_p}/\langle N_p\rangle$ and $\sigma_n/\langle\rho_n\rangle=\sqrt{S_n}/\langle N_n\rangle$. In AMPT, nucleons are extracted for different rapidity ranges and centralities. The definition of centrality is determined by the per-event charged particle multiplicity $N_{\textrm{ch}}$ for pseudorapidity range $\eta\leqslant$ 0.5.

\begin{figure}[tb]
    \includegraphics[width=0.8\textwidth]{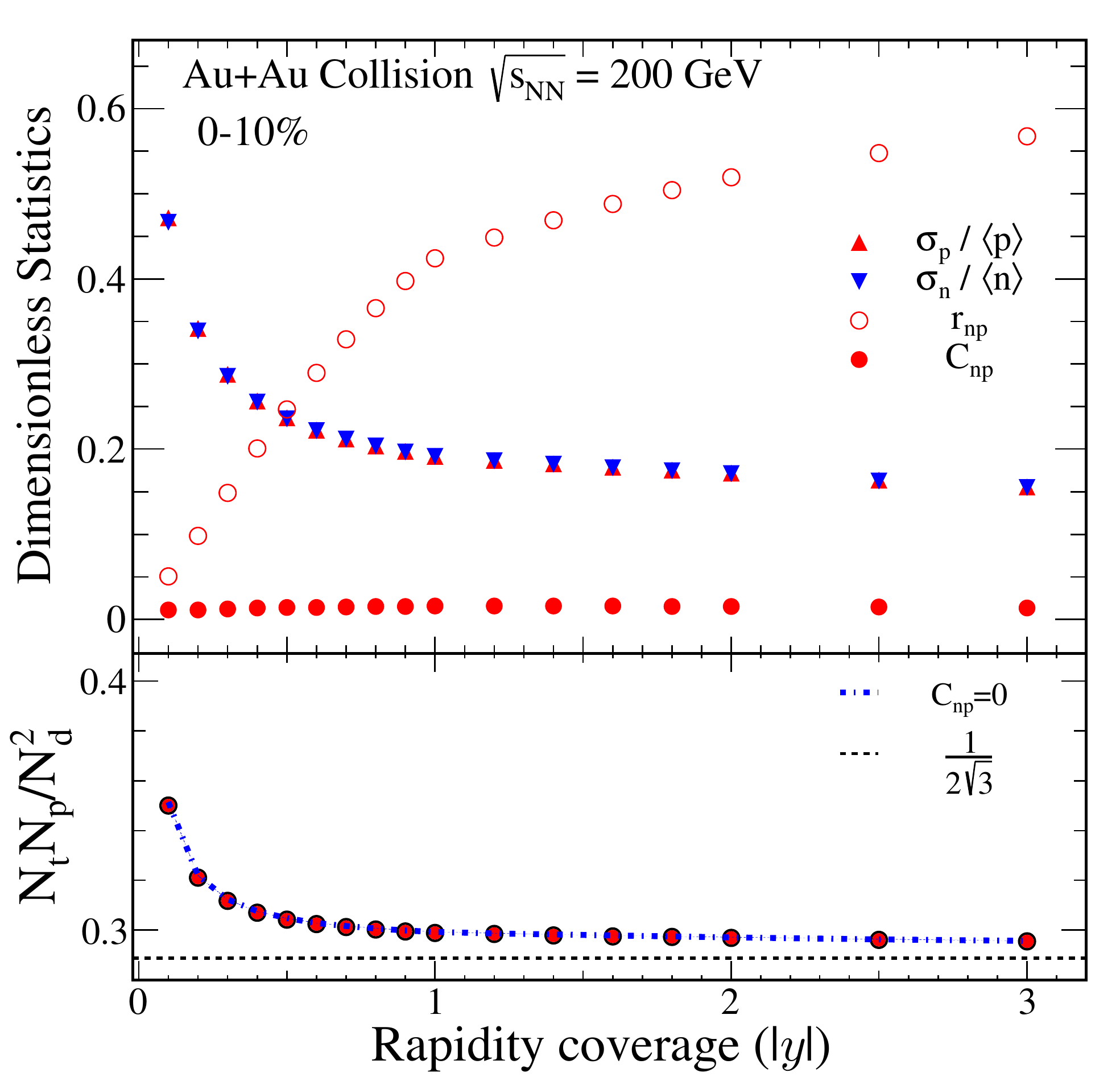}
    \caption{\label{fig1} Top panel: Dimensionless statistics $\sigma_p/\langle p \rangle$, $\sigma_p/\langle p \rangle$, $r_{np}$, and $C_{np}$ for $0-10\%$ Au+Au collisions at $\snn=200$ GeV. Bottom panel: The light nuclei yield ratio $N_tN_p/N_d^2$ calculated from top panel is shown as solid circles by Eq.~(\ref{lnr}) and dash-dot line by Eq.~(\ref{lnr0}).}
\end{figure}

The top panel of Fig.~\ref{fig1} shows the rapidity dependence of the dimensionless statistics $\sigma_p/\langle\rho_p\rangle$, $\sigma_n/\langle\rho_n\rangle$, and $r_{np}$ for $0-10\%$ Au+Au collisions at $\snn=200$ GeV. $\sigma_p/\langle\rho_p\rangle$ and $\sigma_n/\langle\rho_n\rangle$, which can be regarded as relative nucleon density fluctuations, decrease with increasing rapidity coverage. The squares of these quantities are defined as fluctuations in Ref.~\cite{RN9}. In small rapidity coverage, also known as mid-rapidity, pair production is dominant. As a result, fluctuation will be larger at mid-rapidity. As rapidity coverage increases, more and more transport nucleon are taken into account. The total number of transport nucleon is fixed in the collision. This fixed number dilutes the nucleon density fluctuation. Eventually, as rapidity coverage increases, this fluctuation will converge to a constant. $\sigma_p/\langle\rho_p\rangle$ and $\sigma_n/\langle\rho_n\rangle$ are consistent. The correlation coefficient $r_{np}$ reflecting the relationship between neutron and proton density, is positive. That is why $\partial R/\partial C_{np}$ is negative and $R$ decrease with $C_{np}$. At mid-rapidity, there is less correlation between produced protons and neutrons. With the increase of rapidity coverage, the correlation becomes strong and coverage to 0.6 in central collision. The correlation $C_{np}$, which is the product of $\sigma_p/\langle\rho_p\rangle$, $\sigma_n/\langle\rho_n\rangle$, and $r_{np}$, is independent of rapidity coverage and almost vanished for $0-10\%$ Au+Au collisions at $\snn=200$ GeV. From Eq.~(\ref{lnr}), the light nuclei ratio $N_tN_p/N_d^2$ can be calculated. The rapidity dependence of this ratio, the solid circles, is shown in the bottom panel of Fig.~\ref{fig1}. If we assume $C_{np}$ is zero, then Eq.~(\ref{lnr}) can be rewritten as
\begin{equation}
    \frac{N_tN_p}{N_d^2}=\frac{1+\Delta\rho_n}{2\sqrt{3}},
    \label{lnr0}
\end{equation}

The results are shown as dash-dot line in the bottom panel of Fig.~\ref{fig1}. Since $C_{np}$ has all but disappeared in central collision, these two results are consistent. Results from other collision energies at central collision are similar, meaning we are able to extract $\Delta\rho_n$ directly from light nuclei ratio without $C_{np}$ information. We also plot the line of $1/2\sqrt{3}$, which means both density fluctuation and correlation disappeared. It is worth noting that unlike in Ref.~\cite{RN213}, light nuclei ratio is calculated by light nuclei yield, whereas the results in our figures are calculated directly by $\Delta\rho_n$ and $C_{np}$ from AMPT.

\begin{figure}[tb]
    \includegraphics[width=0.8\textwidth]{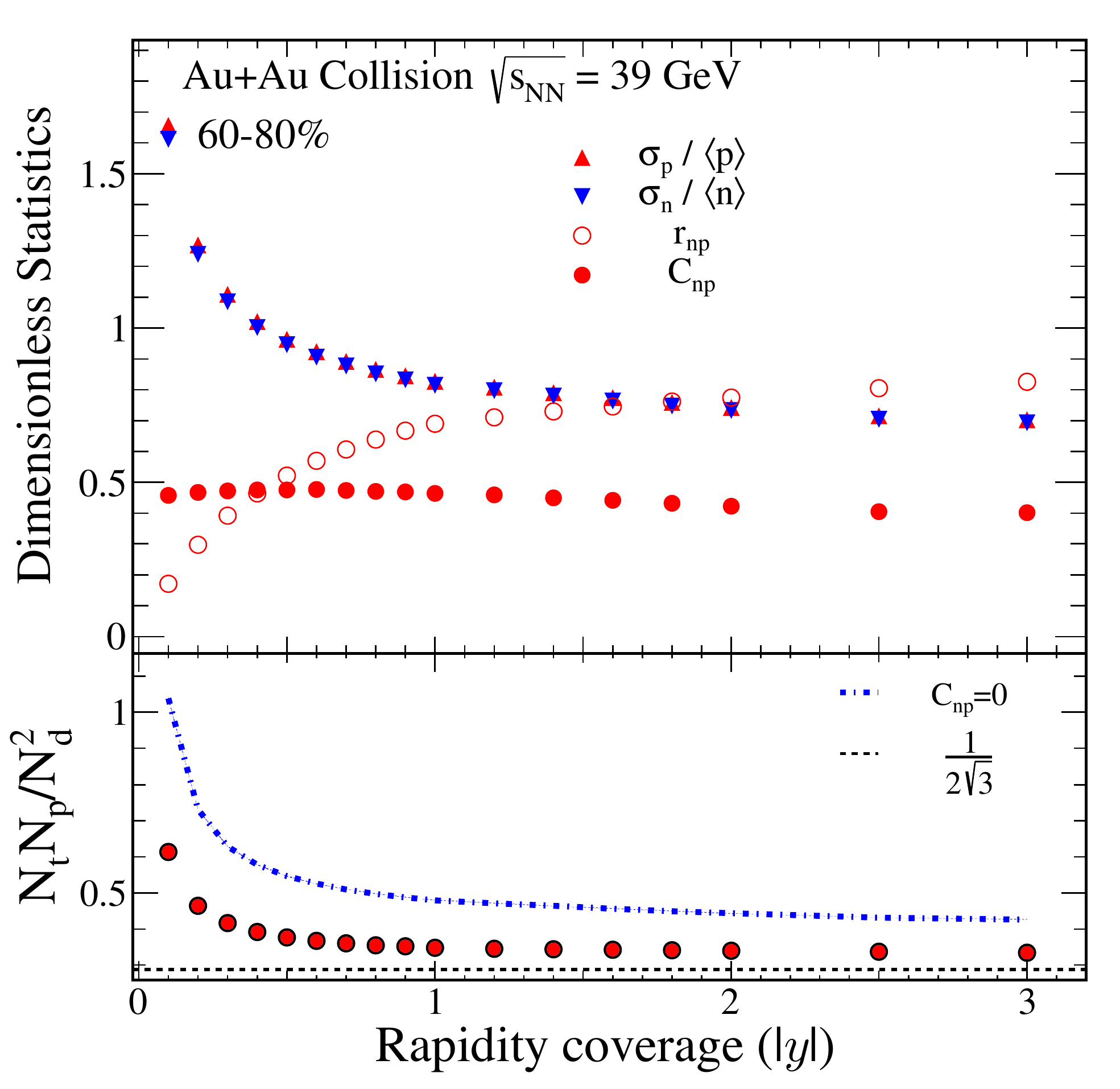}
    \caption{\label{fig2} Top panel: Dimensionless statistics $\sigma_p/\langle p \rangle$, $\sigma_p/\langle p \rangle$, $r_{np}$, and $C_{np}$ for $60-80\%$ Au+Au collisions at $\snn=39$ GeV. Bottom panel: The light nuclei yield ratio $N_tN_p/N_d^2$ calculated from top panel is shown as solid circles by Eq.~(\ref{lnr}) and dash-dot line by Eq.~(\ref{lnr0}).}
\end{figure}

Figure.~\ref{fig2} shows the rapidity dependence of $\sigma_p/\langle\rho_p\rangle$, $\sigma_n/\langle\rho_n\rangle$, and $r_{np}$ for $60-80\%$ Au+Au collisions at $\snn=39$ GeV in the top panel and related light nuclei ratio in the bottom panel. Similar to the results of the central collision, $\sigma_p/\langle\rho_p\rangle$, $\sigma_n/\langle\rho_n\rangle$ decrease and $r_{np}$ increase with increasing rapidity coverage. In a given rapidity coverage, all these three quantities in peripheral collision are larger than those in central collision. The convergence value of $r_{np}$ is about 0.85, which is larger than the convergence value of central collision. The presence of more transport nucleons from the initial collider nuclei in the peripheral collisions may be the main cause. The correlation $C_{np}$ is about $0.45-0.5$, almost independent of rapidity coverage. As a result, the $C_{np}$ can not be neglected in the peripheral collision to calculate the relative neutron density fluctuation from the light nuclei yield ratio. At the bottom of Fig.~\ref{fig2}, we plot the light nuclei yield ratios using Eq.~(\ref{lnr}) and Eq.~(\ref{lnr0}). The results show that neglecting $C_{np}$ leads to a larger light nuclei yield ratio. In other words, if we do not take into account the $C_{np}$, the light nuclei yield ratio will give a smaller relative neutron density fluctuation.

\begin{figure}[tb]
    \includegraphics[width=0.8\textwidth]{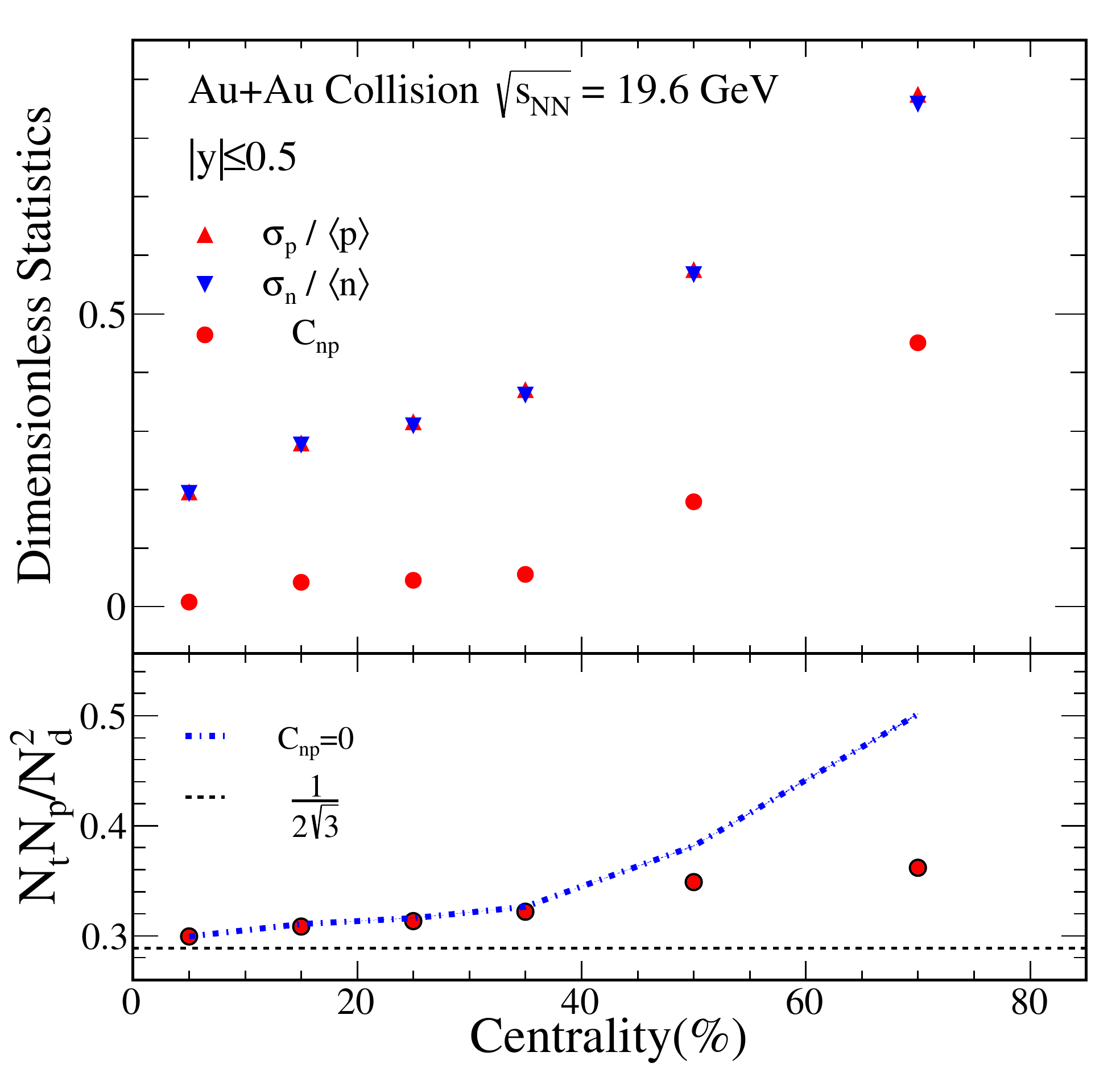}
    \caption{\label{fig3} Top panel: Dimensionless statistics $\sigma_p/\langle p \rangle$, $\sigma_p/\langle p \rangle$, and $C_{np}$ for Au+Au collisions at $\snn=19.6$ GeV with $|y|\leqslant 0.5$. Bottom panel: The light nuclei yield ratio $N_tN_p/N_d^2$ calculated from top panel is shown as solid circles by Eq.~(\ref{lnr}) and dash-dot line by Eq.~(\ref{lnr0}).}
\end{figure}

The centrality dependent $\sigma_p/\langle\rho_p\rangle$, $\sigma_n/\langle\rho_n\rangle$, and $r_{np}$ for Au+Au collisions at $\snn=19.6$ GeV with rapidity coverage of $|y|\leqslant 0.5$ are shown in the top panel of Fig.~\ref{fig3}. All three quantities increase from central to peripheral collision. Smaller relative nucleon density fluctuation implies the nucleon is more evenly distributed at central collision. One possible reason is that quark hadronization creates nucleons that can be produced one by one at different points in phase space. The nucleon density at the point where nucleons are produced is larger than that at the point where nucleons are not produced. The large system size at central collision suppresses this one-by-one effect, while the small system size at peripheral collision enhances it. At the bottom of Fig.~\ref{fig3}, the light nuclei yield ratio corresponds to Eq.~(\ref{lnr}) and Eq.~(\ref{lnr0}) are shown. It can be found that the difference between these two ratios by Eq.~(\ref{lnr}) and Eq.~(\ref{lnr0}) is very small at central or mid-central collision. The relative nucleon density fluctuation can be extracted directly from the light nuclei ratio. In contrast, the effects of neutron-proton correlation $C_{np}$ must be considered in peripheral collisions.

\begin{figure}[tb]
    \includegraphics[width=0.8\textwidth]{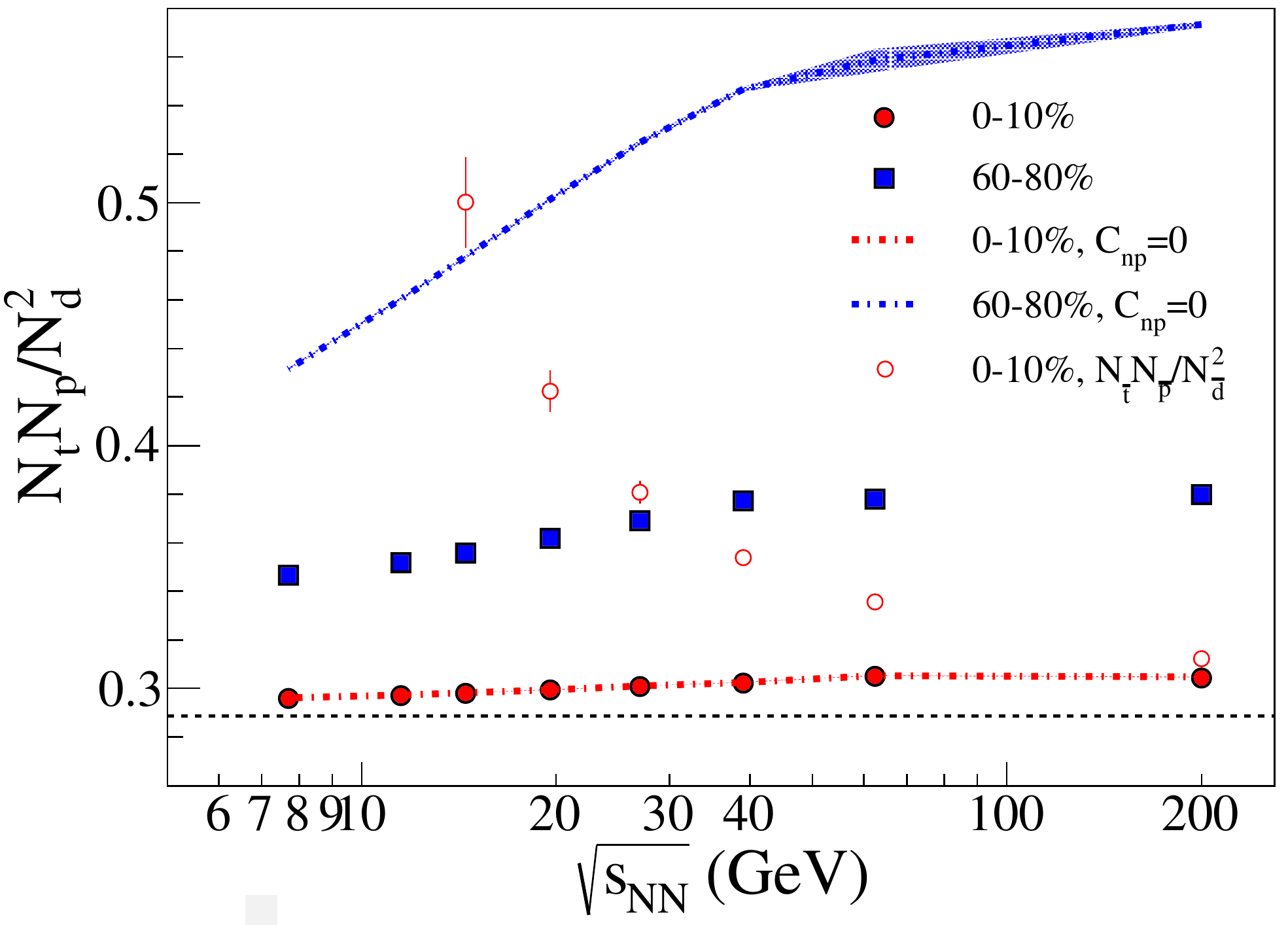}
    \caption{\label{fig4} Collision energy dependence of the light nuclei yield ratio $N_tN_p/N_d^2$ from AMPT for Au+Au collisions with $|y|\leqslant 0.5$. Solid circles and squares are the results of 0\%-10\% central and 60\%-80\% peripheral Au+Au collisions. The dash-dot lines are corresponding results with vanished $C_{np}$. The open circles are the anti-nuclei yield ratio $N_{\bar{t}}N_{\bar{p}}/N_{\bar{d}}^2$ from 0\%-10\% central Au+Au collisions.}
\end{figure}

Figure~\ref{fig4} shows the collision energy dependence of the light nuclei yield ratio $N_tN_p/N_d^2$ from 0\%-10\% central and 60\%-80\% peripheral Au+Au collisions with $|y|\leqslant 0.5$. The results of vanished $C_{np}$ are also shown as dash-dot line. It is clear from the figure that the light nuclei yield ratio increases slightly with increasing collision energy from AMPT model. At 0\%-10\% central Au+Au collisions, the yield ratio is consistent with the results of the coalescence model with vanished relative neutron density fluctuation and $C_{np}$. Because of the system size effect, the yield ratio of peripheral collisions is greater than that of central collisions. The anti-nuclei yield ratio $N_{\bar{t}}N_{\bar{p}}/N_{\bar{d}}^2$ from 0\%-10\% central Au+Au collisions are also shown in Fig.~\ref{fig4}. For certain collision energies and collision centralities, it is assumed that nucleons and anti-nucleons have the same effect volume. The yield of anti-nucleons is smaller than that of nucleons, especially at low energies. In other words, the effective distance between anti-nucleons is less than the effective distance between nucleons. This leads to a smaller $C_{\bar{n}\bar{p}}$. On the other hand, smaller nucleon densities also lead to larger relative neutron density fluctuations $\Delta\rho_{\bar{n}}$. The consequence of smaller $C_{\bar{n}\bar{p}}$ and larger $\Delta\rho_{\bar{n}}$ are larger nuclei yield ratio.

 \begin{figure}[tb]
    \includegraphics[width=0.8\textwidth]{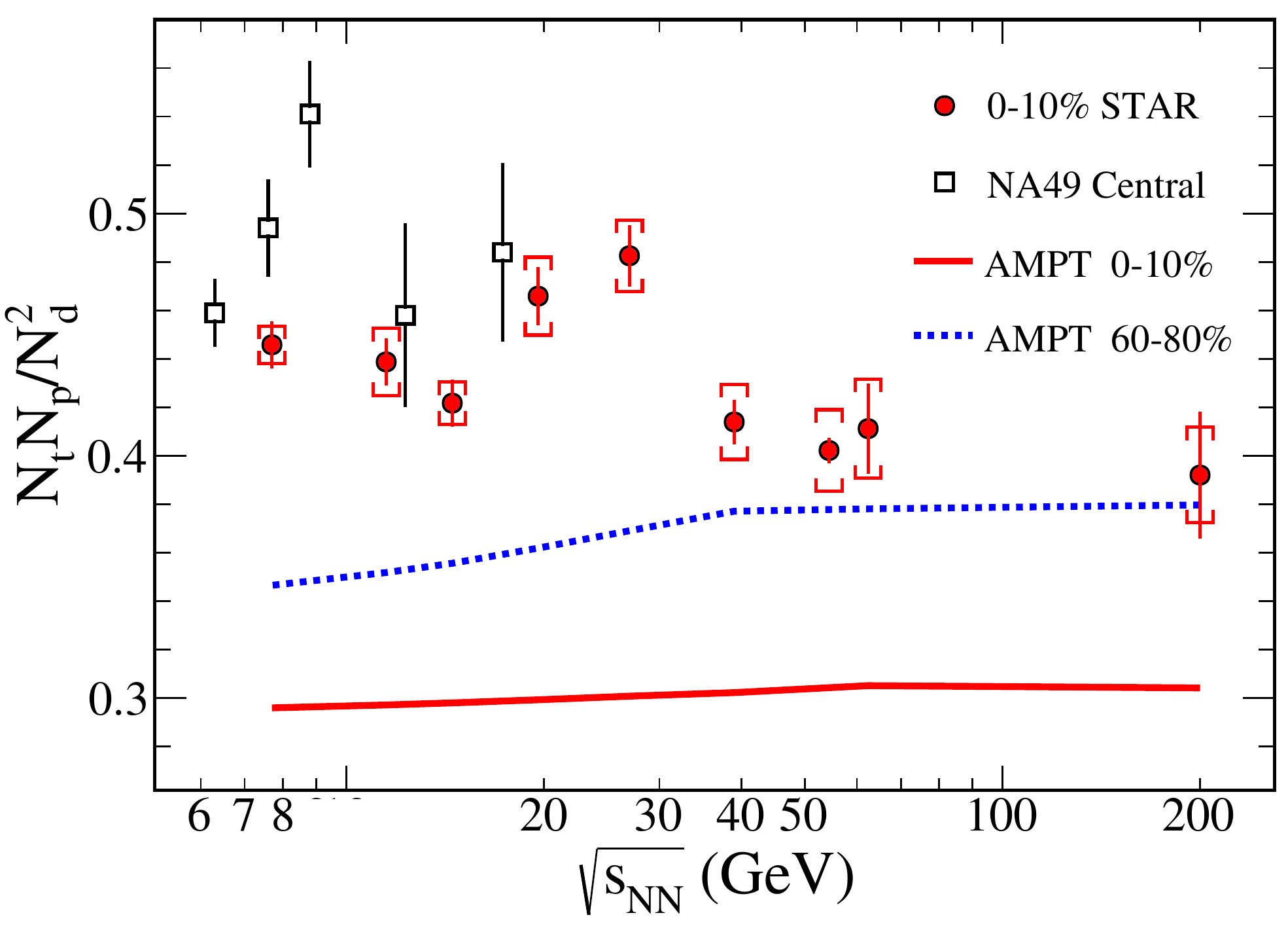}
    \caption{\label{fig5} Collision energy and centrality dependence of the light nuclei yield ratio $N_tN_p/N_d^2$ from AMPT with $|y|\leqslant 0.5$. Solid circles are the results from STAR detector at 0\%-10\% central Au+Au collision~\cite{RN173,2209.08058}. Open squares are the results from NA49 at central Pb+Pb collision~\cite{RN19,RN9}.}
\end{figure}

Figure~\ref{fig5} shows the experimental results of the STAR detector at 0\%-10\% central Au+Au collision~\cite{RN173,2209.08058} and NA49 at central Pb+Pb collision~\cite{RN19,RN9} and compares them with the AMPT results. We can clearly see a non-monotonic energy dependence. The light nuclei yield ratio peaks around $\snn=20\sim30$ GeV, indicating the strongest relative neutron density fluctuation in this energy region. Since the AMPT model does not consider critical physics, it can not describe this non-monotonic energy dependence. It can be used as a baseline without critical physics.

In summary, we investigate the rapidity, collision energy, and centrality dependence of the relative neutron density fluctuation, the neutron-proton correlation $C_{np}$ and the corresponding light nuclei yield ratio $N_{t}N_{p}/N_{d}^2$ by the AMPT model. It is shown that $N_{t}N_{p}/N_{d}^2$ is positively correlated with the relative neutron density fluctuation and negatively correlated with $C_{np}$. At central or mid-central collisions, the correlation $C_{np}$ has little effect on the light nuclei yield ratio $N_{t}N_{p}/N_{d}^2$. In other words, we can extract the relative neutron density fluctuation directly from light nuclei yield ratio. While at peripheral collision, the effect of $C_{np}$ on the light nuclei yield ratio becomes larger, and the related effect must be taken into account when extracting the density fluctuation. From experimental results, the light nuclei yield ratio peaks around $\snn=20\sim30$ GeV, indicating strong relative neutron density fluctuation. Since there is no critical physics in the AMPT model, it underestimates these non-monotonic energy dependence measurements, but it can serve as a baseline for the absence of critical physics.

The authors appreciate the referee for his/her careful reading of the paper and valuable comments. This work is supported in part by the Science and technology research project of Hubei Provincial Department of Education (No.B2021256), Natural Science
Foundation of Henan Province (No.212300410386), Key Research Projects of Henan Higher Education Institutions (No.20A140024), and
NSFC Key Grant 12061141008.
\bibliography{apssamp}

\providecommand{\noopsort}[1]{}\providecommand{\singleletter}[1]{#1}%
\begin{thebibliography}{27}%
\makeatletter
\providecommand \@ifxundefined [1]{%
 \@ifx{#1\undefined}
}%
\providecommand \@ifnum [1]{%
 \ifnum #1\expandafter \@firstoftwo
 \else \expandafter \@secondoftwo
 \fi
}%
\providecommand \@ifx [1]{%
 \ifx #1\expandafter \@firstoftwo
 \else \expandafter \@secondoftwo
 \fi
}%
\providecommand \natexlab [1]{#1}%
\providecommand \enquote  [1]{``#1''}%
\providecommand \bibnamefont  [1]{#1}%
\providecommand \bibfnamefont [1]{#1}%
\providecommand \citenamefont [1]{#1}%
\providecommand \href@noop [0]{\@secondoftwo}%
\providecommand \href [0]{\begingroup \@sanitize@url \@href}%
\providecommand \@href[1]{\@@startlink{#1}\@@href}%
\providecommand \@@href[1]{\endgroup#1\@@endlink}%
\providecommand \@sanitize@url [0]{\catcode `\\12\catcode `\$12\catcode
  `\&12\catcode `\#12\catcode `\^12\catcode `\_12\catcode `\%12\relax}%
\providecommand \@@startlink[1]{}%
\providecommand \@@endlink[0]{}%
\providecommand \url  [0]{\begingroup\@sanitize@url \@url }%
\providecommand \@url [1]{\endgroup\@href {#1}{\urlprefix }}%
\providecommand \urlprefix  [0]{URL }%
\providecommand \Eprint [0]{\href }%
\providecommand \doibase [0]{http://dx.doi.org/}%
\providecommand \selectlanguage [0]{\@gobble}%
\providecommand \bibinfo  [0]{\@secondoftwo}%
\providecommand \bibfield  [0]{\@secondoftwo}%
\providecommand \translation [1]{[#1]}%
\providecommand \BibitemOpen [0]{}%
\providecommand \bibitemStop [0]{}%
\providecommand \bibitemNoStop [0]{.\EOS\space}%
\providecommand \EOS [0]{\spacefactor3000\relax}%
\providecommand \BibitemShut  [1]{\csname bibitem#1\endcsname}%
\let\auto@bib@innerbib\@empty
\bibitem [{\citenamefont {Braun-Munzinger}\ and\ \citenamefont
  {Wambach}(2009)}]{RN21}%
  \BibitemOpen
  \bibfield  {author} {\bibinfo {author} {\bibfnamefont {P.}~\bibnamefont
  {Braun-Munzinger}}\ and\ \bibinfo {author} {\bibfnamefont {J.}~\bibnamefont
  {Wambach}},\ }\href {\doibase 10.1103/RevModPhys.81.1031} {\bibfield
  {journal} {\bibinfo  {journal} {Rev. Mod. Phys.}\ }\textbf {\bibinfo {volume}
  {81}},\ \bibinfo {pages} {1031} (\bibinfo {year} {2009})}\BibitemShut
  {NoStop}%
\bibitem [{\citenamefont {Aoki}\ \emph {et~al.}(2006)\citenamefont {Aoki},
  \citenamefont {Endrodi}, \citenamefont {Fodor}, \citenamefont {Katz},\ and\
  \citenamefont {Szabo}}]{RN175}%
  \BibitemOpen
  \bibfield  {author} {\bibinfo {author} {\bibfnamefont {Y.}~\bibnamefont
  {Aoki}}, \bibinfo {author} {\bibfnamefont {G.}~\bibnamefont {Endrodi}},
  \bibinfo {author} {\bibfnamefont {Z.}~\bibnamefont {Fodor}}, \bibinfo
  {author} {\bibfnamefont {S.~D.}\ \bibnamefont {Katz}}, \ and\ \bibinfo
  {author} {\bibfnamefont {K.~K.}\ \bibnamefont {Szabo}},\ }\href {\doibase
  10.1038/nature05120} {\bibfield  {journal} {\bibinfo  {journal} {Nature}\
  }\textbf {\bibinfo {volume} {443}},\ \bibinfo {pages} {675} (\bibinfo {year}
  {2006})}\BibitemShut {NoStop}%
\bibitem [{\citenamefont {de Forcrand}\ and\ \citenamefont
  {Philipsen}(2002)}]{RN176}%
  \BibitemOpen
  \bibfield  {author} {\bibinfo {author} {\bibfnamefont {P.}~\bibnamefont
  {de Forcrand}}\ and\ \bibinfo {author} {\bibfnamefont {O.}~\bibnamefont
  {Philipsen}},\ }\href {\doibase 10.1016/s0550-3213(02)00626-0} {\bibfield
  {journal} {\bibinfo  {journal} {Nuclear Physics B}\ }\textbf {\bibinfo
  {volume} {642}},\ \bibinfo {pages} {290} (\bibinfo {year}
  {2002})}\BibitemShut {NoStop}%
\bibitem [{\citenamefont {Fodor}\ \emph {et~al.}(2003)\citenamefont {Fodor},
  \citenamefont {Katz},\ and\ \citenamefont {Szabó}}]{RN177}%
  \BibitemOpen
  \bibfield  {author} {\bibinfo {author} {\bibfnamefont {Z.}~\bibnamefont
  {Fodor}}, \bibinfo {author} {\bibfnamefont {S.~D.}\ \bibnamefont {Katz}}, \
  and\ \bibinfo {author} {\bibfnamefont {K.~K.}\ \bibnamefont {Szabó}},\
  }\href {\doibase 10.1016/j.physletb.2003.06.011} {\bibfield  {journal}
  {\bibinfo  {journal} {Phys. Lett. B}\ }\textbf {\bibinfo {volume} {568}},\
  \bibinfo {pages} {73} (\bibinfo {year} {2003})}\BibitemShut {NoStop}%
\bibitem [{\citenamefont {Fodor}\ and\ \citenamefont {Katz}(2004)}]{RN178}%
  \BibitemOpen
  \bibfield  {author} {\bibinfo {author} {\bibfnamefont {Z.}~\bibnamefont
  {Fodor}}\ and\ \bibinfo {author} {\bibfnamefont {S.~D.}\ \bibnamefont
  {Katz}},\ }\href {\doibase 10.1088/1126-6708/2004/04/050} {\bibfield
  {journal} {\bibinfo  {journal} {J. High Energy Phys.}\ }\textbf {\bibinfo
  {volume} {2004}},\ \bibinfo {pages} {050} (\bibinfo {year}
  {2004})}\BibitemShut {NoStop}%
\bibitem [{\citenamefont {Gavai}\ and\ \citenamefont {Gupta}(2005)}]{RN179}%
  \BibitemOpen
  \bibfield  {author} {\bibinfo {author} {\bibfnamefont {R.~V.}\ \bibnamefont
  {Gavai}}\ and\ \bibinfo {author} {\bibfnamefont {S.}~\bibnamefont {Gupta}},\
  }\href {\doibase 10.1103/PhysRevD.71.114014} {\bibfield  {journal} {\bibinfo
  {journal} {Phys. Rev. D}\ }\textbf {\bibinfo {volume} {71}} (\bibinfo {year}
  {2005}),\ 10.1103/PhysRevD.71.114014}\BibitemShut {NoStop}%
\bibitem [{\citenamefont {Karsch}\ \emph {et~al.}(2016)\citenamefont {Karsch}
  \emph {et~al.}}]{RN180}%
  \BibitemOpen
  \bibfield  {author} {\bibinfo {author} {\bibfnamefont {F.}~\bibnamefont
  {Karsch}} \emph {et~al.},\ }\href {\doibase 10.1016/j.nuclphysa.2016.01.008}
  {\bibfield  {journal} {\bibinfo  {journal} {Nucl. Phys. A}\ }\textbf
  {\bibinfo {volume} {956}},\ \bibinfo {pages} {352} (\bibinfo {year}
  {2016})}\BibitemShut {NoStop}%
\bibitem [{\citenamefont {Gupta}\ \emph {et~al.}(2011)\citenamefont {Gupta},
  \citenamefont {Luo}, \citenamefont {Mohanty}, \citenamefont {Ritter},\ and\
  \citenamefont {Xu}}]{RN181}%
  \BibitemOpen
  \bibfield  {author} {\bibinfo {author} {\bibfnamefont {S.}~\bibnamefont
  {Gupta}}, \bibinfo {author} {\bibfnamefont {X.}~\bibnamefont {Luo}}, \bibinfo
  {author} {\bibfnamefont {B.}~\bibnamefont {Mohanty}}, \bibinfo {author}
  {\bibfnamefont {H.~G.}\ \bibnamefont {Ritter}}, \ and\ \bibinfo {author}
  {\bibfnamefont {N.}~\bibnamefont {Xu}},\ }\href {\doibase
  10.1126/science.1204621} {\bibfield  {journal} {\bibinfo  {journal}
  {Science}\ }\textbf {\bibinfo {volume} {332}},\ \bibinfo {pages} {1525}
  (\bibinfo {year} {2011})}\BibitemShut {NoStop}%
\bibitem [{\citenamefont {Stephanov}(2006)}]{RN182}%
  \BibitemOpen
  \bibfield  {author} {\bibinfo {author} {\bibfnamefont {M.}~\bibnamefont
  {Stephanov}},\ }\href {\doibase 10.22323/1.032.0024} {\bibfield  {journal}
  {\bibinfo  {journal} {PoS}\ }\textbf {\bibinfo {volume} {LAT2006}},\ \bibinfo
  {pages} {024} (\bibinfo {year} {2006})}\BibitemShut {NoStop}%
\bibitem [{\citenamefont {Aggarwal}\ \emph {et~al.}(2010)\citenamefont
  {Aggarwal} \emph {et~al.}}]{RN183}%
  \BibitemOpen
  \bibfield  {author} {\bibinfo {author} {\bibfnamefont {M.~M.}\ \bibnamefont
  {Aggarwal}} \emph {et~al.},\ }\href {\doibase 10.1103/PhysRevLett.105.022302}
  {\bibfield  {journal} {\bibinfo  {journal} {Phys Rev Lett}\ }\textbf
  {\bibinfo {volume} {105}},\ \bibinfo {pages} {022302} (\bibinfo {year}
  {2010})}\BibitemShut {NoStop}%
\bibitem [{\citenamefont {Adamczyk}\ \emph
  {et~al.}(2014{\natexlab{a}})\citenamefont {Adamczyk} \emph {et~al.}}]{RN184}%
  \BibitemOpen
  \bibfield  {author} {\bibinfo {author} {\bibfnamefont {L.}~\bibnamefont
  {Adamczyk}} \emph {et~al.},\ }\href {\doibase 10.1103/PhysRevLett.112.032302}
  {\bibfield  {journal} {\bibinfo  {journal} {Phys Rev Lett}\ }\textbf
  {\bibinfo {volume} {112}},\ \bibinfo {pages} {032302} (\bibinfo {year}
  {2014}{\natexlab{a}})}\BibitemShut {NoStop}%
\bibitem [{\citenamefont {Adamczyk}\ \emph
  {et~al.}(2014{\natexlab{b}})\citenamefont {Adamczyk} \emph {et~al.}}]{RN185}%
  \BibitemOpen
  \bibfield  {author} {\bibinfo {author} {\bibfnamefont {L.}~\bibnamefont
  {Adamczyk}} \emph {et~al.},\ }\href {\doibase 10.1103/PhysRevLett.113.092301}
  {\bibfield  {journal} {\bibinfo  {journal} {Phys Rev Lett}\ }\textbf
  {\bibinfo {volume} {113}},\ \bibinfo {pages} {092301} (\bibinfo {year}
  {2014}{\natexlab{b}})}\BibitemShut {NoStop}%
\bibitem [{\citenamefont {Adamczyk}\ \emph {et~al.}(2018)\citenamefont
  {Adamczyk} \emph {et~al.}}]{RN186}%
  \BibitemOpen
  \bibfield  {author} {\bibinfo {author} {\bibfnamefont {L.}~\bibnamefont
  {Adamczyk}} \emph {et~al.},\ }\href {\doibase 10.1016/j.physletb.2018.07.066}
  {\bibfield  {journal} {\bibinfo  {journal} {Phys. Lett. B}\ }\textbf
  {\bibinfo {volume} {785}},\ \bibinfo {pages} {551} (\bibinfo {year}
  {2018})}\BibitemShut {NoStop}%
\bibitem [{\citenamefont {Adam}\ \emph {et~al.}(2019)\citenamefont {Adam} \emph
  {et~al.}}]{RN187}%
  \BibitemOpen
  \bibfield  {author} {\bibinfo {author} {\bibfnamefont {J.}~\bibnamefont
  {Adam}} \emph {et~al.},\ }\href {\doibase 10.1103/PhysRevC.100.014902}
  {\bibfield  {journal} {\bibinfo  {journal} {Physical Review C}\ }\textbf
  {\bibinfo {volume} {100}} (\bibinfo {year} {2019}),\
  10.1103/PhysRevC.100.014902}\BibitemShut {NoStop}%
\bibitem [{\citenamefont {Adam}\ \emph {et~al.}(2021)\citenamefont {Adam} \emph
  {et~al.}}]{RN188}%
  \BibitemOpen
  \bibfield  {author} {\bibinfo {author} {\bibfnamefont {J.}~\bibnamefont
  {Adam}} \emph {et~al.},\ }\href {\doibase 10.1103/PhysRevLett.126.092301}
  {\bibfield  {journal} {\bibinfo  {journal} {Phys Rev Lett}\ }\textbf
  {\bibinfo {volume} {126}},\ \bibinfo {pages} {092301} (\bibinfo {year}
  {2021})}\BibitemShut {NoStop}%
\bibitem [{\citenamefont {Sun}\ \emph {et~al.}(2017)\citenamefont {Sun},
  \citenamefont {Chen}, \citenamefont {Ko},\ and\ \citenamefont {Xu}}]{RN14}%
  \BibitemOpen
  \bibfield  {author} {\bibinfo {author} {\bibfnamefont {K.-J.}\ \bibnamefont
  {Sun}}, \bibinfo {author} {\bibfnamefont {L.-W.}\ \bibnamefont {Chen}},
  \bibinfo {author} {\bibfnamefont {C.~M.}\ \bibnamefont {Ko}}, \ and\ \bibinfo
  {author} {\bibfnamefont {Z.}~\bibnamefont {Xu}},\ }\href {\doibase
  10.1016/j.physletb.2017.09.056} {\bibfield  {journal} {\bibinfo  {journal}
  {Physics Letters B}\ }\textbf {\bibinfo {volume} {774}},\ \bibinfo {pages}
  {103} (\bibinfo {year} {2017})}\BibitemShut {NoStop}%
\bibitem [{\citenamefont {Sun}\ \emph {et~al.}(2018)\citenamefont {Sun},
  \citenamefont {Chen}, \citenamefont {Ko}, \citenamefont {Pu},\ and\
  \citenamefont {Xu}}]{RN9}%
  \BibitemOpen
  \bibfield  {author} {\bibinfo {author} {\bibfnamefont {K.-J.}\ \bibnamefont
  {Sun}}, \bibinfo {author} {\bibfnamefont {L.-W.}\ \bibnamefont {Chen}},
  \bibinfo {author} {\bibfnamefont {C.~M.}\ \bibnamefont {Ko}}, \bibinfo
  {author} {\bibfnamefont {J.}~\bibnamefont {Pu}}, \ and\ \bibinfo {author}
  {\bibfnamefont {Z.}~\bibnamefont {Xu}},\ }\href {\doibase
  10.1016/j.physletb.2018.04.035} {\bibfield  {journal} {\bibinfo  {journal}
  {Phys. Lett. B}\ }\textbf {\bibinfo {volume} {781}},\ \bibinfo {pages} {499}
  (\bibinfo {year} {2018})}\BibitemShut {NoStop}%
\bibitem [{\citenamefont {Yu}\ \emph {et~al.}(2020)\citenamefont {Yu},
  \citenamefont {Zhang},\ and\ \citenamefont {Luo}}]{RN118}%
  \BibitemOpen
  \bibfield  {author} {\bibinfo {author} {\bibfnamefont {N.}~\bibnamefont
  {Yu}}, \bibinfo {author} {\bibfnamefont {D.}~\bibnamefont {Zhang}}, \ and\
  \bibinfo {author} {\bibfnamefont {X.}~\bibnamefont {Luo}},\ }\href {\doibase
  10.1088/1674-1137/44/1/014002} {\bibfield  {journal} {\bibinfo  {journal}
  {Chin. Phys. C}\ }\textbf {\bibinfo {volume} {44}} (\bibinfo {year} {2020}),\
  10.1088/1674-1137/44/1/014002}\BibitemShut {NoStop}%
\bibitem [{\citenamefont {Zhao}\ \emph {et~al.}(2021)\citenamefont {Zhao},
  \citenamefont {Sun}, \citenamefont {Ko},\ and\ \citenamefont {Luo}}]{RN216}%
  \BibitemOpen
  \bibfield  {author} {\bibinfo {author} {\bibfnamefont {W.}~\bibnamefont
  {Zhao}}, \bibinfo {author} {\bibfnamefont {K.-j.}\ \bibnamefont {Sun}},
  \bibinfo {author} {\bibfnamefont {C.~M.}\ \bibnamefont {Ko}}, \ and\ \bibinfo
  {author} {\bibfnamefont {X.}~\bibnamefont {Luo}},\ }\href {\doibase
  10.1016/j.physletb.2021.136571} {\bibfield  {journal} {\bibinfo  {journal}
  {Physics Letters B}\ }\textbf {\bibinfo {volume} {820}} (\bibinfo {year}
  {2021}),\ 10.1016/j.physletb.2021.136571}\BibitemShut {NoStop}%
\bibitem [{\citenamefont {Liu}\ \emph {et~al.}(2020)\citenamefont {Liu},
  \citenamefont {Zhang}, \citenamefont {He}, \citenamefont {Sun}, \citenamefont
  {Yu},\ and\ \citenamefont {Luo}}]{RN172}%
  \BibitemOpen
  \bibfield  {author} {\bibinfo {author} {\bibfnamefont {H.}~\bibnamefont
  {Liu}}, \bibinfo {author} {\bibfnamefont {D.}~\bibnamefont {Zhang}}, \bibinfo
  {author} {\bibfnamefont {S.}~\bibnamefont {He}}, \bibinfo {author}
  {\bibfnamefont {K.-j.}\ \bibnamefont {Sun}}, \bibinfo {author} {\bibfnamefont
  {N.}~\bibnamefont {Yu}}, \ and\ \bibinfo {author} {\bibfnamefont
  {X.}~\bibnamefont {Luo}},\ }\href {\doibase 10.1016/j.physletb.2020.135452}
  {\bibfield  {journal} {\bibinfo  {journal} {Phys. Lett. B}\ }\textbf
  {\bibinfo {volume} {805}} (\bibinfo {year} {2020}),\
  10.1016/j.physletb.2020.135452}\BibitemShut {NoStop}%
\bibitem [{\citenamefont {Zhang}(2021)}]{RN173}%
  \BibitemOpen
  \bibfield  {author} {\bibinfo {author} {\bibfnamefont {D.}~\bibnamefont
  {Zhang}},\ }\href {\doibase 10.1016/j.nuclphysa.2020.121825} {\bibfield
  {journal} {\bibinfo  {journal} {Nucl. Phys. A}\ }\textbf {\bibinfo {volume}
  {1005}} (\bibinfo {year} {2021}),\
  10.1016/j.nuclphysa.2020.121825}\BibitemShut {NoStop}%
\bibitem [{\citenamefont {Collaboration}(2022)}]{2209.08058}%
  \BibitemOpen
  \bibfield  {author} {\bibinfo {author} {\bibfnamefont {T.~S.}\ \bibnamefont
  {Collaboration}},\ }\href@noop {} {} (\bibinfo {year} {2022}),\ \Eprint
  {http://arxiv.org/abs/2209.08058} {2209.08058} \BibitemShut {NoStop}%
\bibitem [{\citenamefont {Lin}\ \emph {et~al.}(2005)\citenamefont {Lin},
  \citenamefont {Ko}, \citenamefont {Li}, \citenamefont {Zhang},\ and\
  \citenamefont {Pal}}]{RN47}%
  \BibitemOpen
  \bibfield  {author} {\bibinfo {author} {\bibfnamefont {Z.-W.}\ \bibnamefont
  {Lin}}, \bibinfo {author} {\bibfnamefont {C.~M.}\ \bibnamefont {Ko}},
  \bibinfo {author} {\bibfnamefont {B.-A.}\ \bibnamefont {Li}}, \bibinfo
  {author} {\bibfnamefont {B.}~\bibnamefont {Zhang}}, \ and\ \bibinfo {author}
  {\bibfnamefont {S.}~\bibnamefont {Pal}},\ }\href {\doibase
  10.1103/PhysRevC.72.064901} {\bibfield  {journal} {\bibinfo  {journal}
  {Physical Review C}\ }\textbf {\bibinfo {volume} {72}} (\bibinfo {year}
  {2005}),\ 10.1103/PhysRevC.72.064901}\BibitemShut {NoStop}%
\bibitem [{\citenamefont {Andersson}\ \emph {et~al.}(1983)\citenamefont
  {Andersson}, \citenamefont {Gustafson}, \citenamefont {Ingelman},\ and\
  \citenamefont {Sjöstrand}}]{RN212}%
  \BibitemOpen
  \bibfield  {author} {\bibinfo {author} {\bibfnamefont {B.}~\bibnamefont
  {Andersson}}, \bibinfo {author} {\bibfnamefont {G.}~\bibnamefont
  {Gustafson}}, \bibinfo {author} {\bibfnamefont {G.}~\bibnamefont {Ingelman}},
  \ and\ \bibinfo {author} {\bibfnamefont {T.}~\bibnamefont {Sjöstrand}},\
  }\href {\doibase 10.1016/0370-1573(83)90080-7} {\bibfield  {journal}
  {\bibinfo  {journal} {Physics Reports}\ }\textbf {\bibinfo {volume} {97}},\
  \bibinfo {pages} {31} (\bibinfo {year} {1983})}\BibitemShut {NoStop}%
\bibitem [{\citenamefont {Lin}(2014)}]{RN97}%
  \BibitemOpen
  \bibfield  {author} {\bibinfo {author} {\bibfnamefont {Z.-W.}\ \bibnamefont
  {Lin}},\ }\href {\doibase 10.1103/PhysRevC.90.014904} {\bibfield  {journal}
  {\bibinfo  {journal} {Physical Review C}\ }\textbf {\bibinfo {volume} {90}}
  (\bibinfo {year} {2014}),\ 10.1103/PhysRevC.90.014904}\BibitemShut {NoStop}%
\bibitem [{\citenamefont {Sun}\ \emph {et~al.}(2021)\citenamefont {Sun},
  \citenamefont {Ko},\ and\ \citenamefont {Lin}}]{RN213}%
  \BibitemOpen
  \bibfield  {author} {\bibinfo {author} {\bibfnamefont {K.-J.}\ \bibnamefont
  {Sun}}, \bibinfo {author} {\bibfnamefont {C.~M.}\ \bibnamefont {Ko}}, \ and\
  \bibinfo {author} {\bibfnamefont {Z.-W.}\ \bibnamefont {Lin}},\ }\href
  {\doibase 10.1103/PhysRevC.103.064909} {\bibfield  {journal} {\bibinfo
  {journal} {Physical Review C}\ }\textbf {\bibinfo {volume} {103}} (\bibinfo
  {year} {2021}),\ 10.1103/PhysRevC.103.064909}\BibitemShut {NoStop}%
\bibitem [{\citenamefont {Anticic}\ \emph {et~al.}(2016)\citenamefont {Anticic}
  \emph {et~al.}}]{RN19}%
  \BibitemOpen
  \bibfield  {author} {\bibinfo {author} {\bibfnamefont {T.}~\bibnamefont
  {Anticic}} \emph {et~al.} (\bibinfo {collaboration} {NA49 Collaboration}),\
  }\href {\doibase 10.1103/PhysRevC.94.044906} {\bibfield  {journal} {\bibinfo
  {journal} {Physical Review C}\ }\textbf {\bibinfo {volume} {94}} (\bibinfo
  {year} {2016}),\ 10.1103/PhysRevC.94.044906}\BibitemShut {NoStop}%
\end{thebibliography}%

\end{document}